\documentclass[12pt,tightenlines,eqsecnum,floats,aps,amsmath,amssymb,nofootinbib,prd,superscriptaddress]{revtex4}
\usepackage{setspace}
\usepackage{amsmath,amssymb,amsfonts,amsthm,mathrsfs}
\usepackage{graphicx}
\usepackage{enumerate} 
\usepackage{comment}

\usepackage[pdftex]{hyperref}

\def\be{\begin{equation}}
\def\ee{\end{equation}}
\def\ba{\begin{eqnarray}}
\def\ea{\end{eqnarray}}
\def\bi{\begin{itemize}}
\def\ei{\end{itemize}}

\def\xh{\hat{x}}

\def\w{\omega}
\def\G{\mathcal{G}}
\def\lam{\lambda}

\def\F{\mathcal{F}}
\def\I{\mathcal{I}}
\def\Ip{\mathcal{I}^+}
\def\Fo{\overset{0}{F}}
\def\Fone{\overset{-1}{F}}
\def\Ftwo{\overset{-2}{F}}
\def\Fn{\overset{-n}{F}}
\def\Fk{\overset{-k}{F}}
\def\jn{\overset{-n}{j}}
\def\D{\mathcal{D}}
\def\H{\mathcal{H}^o}
\def\Aone{\overset{-1}{A}}
\def\Ao{\overset{0}{A}}

\def\Aoa{\overset{0}{A_\alpha}}
\def\An{\overset{-n}{A}}
\def\t{\tau}
\def\wbulk{\omega_{\text{bulk}}}
\def\wbdy{\omega_{\text{bdy}}}

\def\Lamn{\overset{-n}{\Lambda}}
\def\Lamln{\overset{-n, \ln}{\Lambda}}

\def\Fso{\overset{0}{F^*}}
\def\Fsone{\overset{-1}{F^*}}
\def\Fstwo{\overset{-2}{F^*}}
\def\e{\epsilon}
\def\a{\alpha}
\def\b{\beta}
\def\g{\gamma}
\def\Hp{\mathcal{H}^{+}}
\def\jthree{\overset{-3}{j}}
\def\psit{\tilde{\psi}}
\def\r{\rho}
\def\Qh{Q_{\Hp}}

\def\Qp{Q^{i^+}}
\def\Qtp{\tilde{Q}^{i^+}}

\def\Qt{\tilde{Q}}
\def\Okg{\Omega^{\text{KG}}}
\def\Lam{\Lambda}

\begin{document}

\title{Asymptotic $U(1)$ charges at spatial infinity}
\author{Miguel Campiglia} \email{campi@fisica.edu.uy}
\affiliation{Instituto de F\'isica, Facultad de Ciencias,  Montevideo 11400, Uruguay}
\author{Rodrigo Eyheralde}\email{reyheralde@fisica.edu.uy}
 \affiliation{Instituto de F\'isica, Facultad de Ciencias,  Montevideo 11400, Uruguay}

\begin{abstract}
Large gauge symmetries in Minkowski spacetime are often  studied  in two distinct regimes: either at asymptotic (past or future)  times  or at spatial infinity. By working in harmonic gauge,  we provide a unified description of  large gauge symmetries (and their associated charges) that applies to both regimes. At spatial infinity the charges are conserved and interpolate between those defined at the asymptotic past and future. This  explains the equality of asymptotic past and future charges, as recently proposed in connection with Weinberg's soft photon theorem. 
\end{abstract}
\maketitle

\tableofcontents
\section{Introduction} \label{intro}

The subject of asymptotic symmetries in gravity and gauge theories has experienced renewed interest  after the discovered \cite{strom0}  relation with so-called soft theorems \cite{weinberg}.  In general relativity, the occurrence  of novel asymptotic symmetries 
originally appeared
in the study of gravitational waves  \cite{bms} whereas in electromagnetism,  large $U(1)$ symmetries were studied  in relation to infrared divergences in QED \cite{gz}. There is in fact a tight connection between  asymptotic symmetries and infrared issues in both  gravity and electromagnetism as emphasized long ago \cite{aabook} and more recently \cite{kf}.



The key insight of Strominger and collaborators was the realization that the well-established soft theorems 
can be understood as the existence of infinitely many conserved charges associated to  asymptotic symmetries. 
Here we would like to better understand  such charges and their conservation from a  classical field theory  perspective.
In the case of electromagnetism, the conservation statement associated to the leading soft photon theorem can be written as the following ``continuity'' condition  of the field strength at future and past null infinities \cite{strommass}:
\be
\Ftwo_{ru}(u=-\infty,\xh) = \Ftwo_{rv}(v=\infty,-\xh). \label{newcons}
\ee
Here $\Ftwo_{ru}(u,\xh)$ is the leading term in an $r \to \infty$ expansion of the field strength at future null infinity and  $\Ftwo_{rv}$ the analogue term  at past null infinity.\footnote{The quantity  $\Ftwo_{ru}(u,\xh)$ is also referred to as `charge aspect' in analogy to Bondi's `mass aspect'. The conservation statement associated to the leading soft graviton theorem is a continuity condition on Bondi's mass aspect  \cite{strom0}. Similar continuity conditions are behind  soft theorems for other spins \cite{onehalf,threehalfs,higherspin,mizera}.}

Our first goal will be to show how  condition (\ref{newcons}) can be understood as a consequence of Maxwell equations near spatial infinity. Here we will follow the idea \cite{hansen,held,tn} of using the space-infinity hyperboloid as a way to link future and past null infinities. 

We will then study large $U(1)$ charges from the perspective of spatial infinity. As reviewed below, the smeared version of the quantities (\ref{newcons}) have the interpretation of generators of large $U(1)$ gauge transformations at future and past infinities. We will use the expansion of the Maxwell field at spatial infinity \cite{hansen,held,beig,bergmann} to define large $U(1)$ generators at \emph{arbitrary times} on the space-infinity hyperboloid. The charges will be shown to be conserved and to interpolate between the future and past infinity charges. 
 The end result will be a unified description of large $U(1)$ symmetries and charges that applies to both spatial and future/past infinities. In this way, the present article extends the work \cite{clmass} that provides a unified picture of null and time-infinities charges.



We would like to mention that 
our first result, namely deriving Eq. (\ref{newcons}) from the field equations (plus fall-off assumptions), is not new but was established by  Herdegen long ago \cite{herdlong}. Herdegen's techniques  are however quite different from the ones presented here and we leave for the future a detailed comparison between the two approaches.\footnote{We take the opportunity to address Herdgen's concerns \cite{herdcrit} regarding the interpretation of the charges as generators of large gauge transformations. Herdegen works in Lorenz gauge  and concludes there are no large residual gauge transformation. 
In our understanding, his conclusion may be reached as follows:  In Lorenz gauge, residual gauge transformations are given by gauge parameters satisfying the wave equation $\square \Lambda=0$. Regular solutions to the wave equation are known to decay as $1/r$ in null  directions (see e.g. \cite{wald}). Hence large gauge transformations are forbidden. Indeed, the large gauge parameters considered here and in  \cite{clmass,subqed} are singular at $r=0$. In our view, the catch in the argument is that in the discussions of large gauge symmetries the gauge conditions are only being used \emph{asymptotically}. One may envisage a modification of the Lorenz gauge condition in the interior so as to ensure a regular gauge parameter everywhere.}

The outline of the paper is as follows. In section \ref{prelsec} we set up notation and describe the assumed asymptotic properties of the fields at null, spatial and time infinities. Some derivations in this section are given in appendix \ref{appA}. In section \ref{sec3} we present the argument for Eq. (\ref{conslaw}). The argument rests on properties of the field equations at spatial infinity that are described in appendix \ref{waveeqdS}. In section \ref{pssec} we study large gauge symmetries at spatial infinity and obtain the associated charges.  The analysis requires a careful handling of the covariant phase space which may be of interest in its own right.  In section \ref{magsec} we describe how dual magnetic charges may be constructed. In the final section we summarize our results and point out potential connections with other works. 

\section{Preliminaries} \label{prelsec}
The system of study will be a Maxwell field $A_a$ in four dimensional flat spacetime coupled to massive charge fields. For definitiveness we consider  a single massive charged scalar $\varphi$ but the discussion applies to  any number and kind of massive particles.  The field equations are
 \ba
\nabla^b F_{a b} & = & j_a ,  \label{eom} \\
\quad \quad  \D^a \D_a \phi -m^2 \phi & = &0, \label{eom2}
\ea
where $\nabla_a$ is the spacetime covariant derivative, $\F_{a b}= \partial_a A_b - \partial_b A_a$ the field strength, 
\be
j_a= i e \phi (\D_a \phi)^* + c.c.,
\ee
 the charge current and $\D_a$ the gauge covariant derivative, $\D_a \phi = \partial_a \phi - i e A_a \phi$.
 
Local $U(1)$ gauge transformations are parametrized by a scalar $\Lam$ and act as
\be
\delta_{\Lam} A_a  =  \partial_a \Lam, \quad \delta_{\Lam} \phi =  i e \Lam \phi \label{gge}.
\ee
Large $U(1)$ gauge parameter will be discussed in the context of Lorenz gauge
\be
\nabla^a A_a =0,
\ee
in which the \emph{residual} gauge parameters $\Lambda$ satisfy the wave equation
\be
\nabla^a \nabla_a \Lambda=0. \label{eom3}
\ee
In the following subsections we describe  expansions of the fields at null, space and time infinities. We will assume that in a neighborhood of each infinity the system is described by the free (linearized) theory. The scalar field being massive will only register at time-like infinity. The Maxwell field will have nontrivial components at all  infinities.\\

\noindent {\bf Notation:}\\
The  discussion involves many spaces and notation may be confusing at first. We  here summarize  the notation for coordinates, metric and derivative for each space:
\bi
\item spacetime $\mathbb{R}^4$: $x^a, \eta_{ab}, \nabla_a$
\item sphere $S^2$: $x^A$, $q_{AB}$, $D_A$
\item spatial infinity $\H$: $y^\alpha=(\t,x^A)$, $h_{\alpha \beta}$, $D_\alpha$
\item future time-like infinity $\Hp$: $y^\alpha=(\rho,x^A)$, $h_{\alpha \beta}$, $D_\alpha$
\item future null infinity $\I^+$: $(u,x^A)$, $q_{AB}$, $(\partial_u,D_A)$
\ei

$\xh$ denotes the unit three-vector determine by a sphere point $x^A$. Functions $f$ on the sphere are  denoted as $f(\xh)$.

We warn the reader that the coordinates $\rho$ and $\t$ have different meaning depending on whether we are discussing expansions at space or at time infinities.

\subsection{Field expansion at null infinity}
The (future) null infinity regime can be studied in outgoing coordinates  in terms of which  the Minkowski line element takes the form
\be
ds^2= - du^2 - 2 du dr +r^2 q_{AB} d x^A d x^B.
\ee
Here $r= \sqrt{x^2+y^2+z^2}$ is the radial coordinate, $u=t-r$ the retarded time, and $x^A,  A=1,2$  coordinates on the sphere.  $q_{AB}$ is the unit sphere metric. We use the notation $\xh$ to denote the unit three-vector parametrized by the point on the sphere $x^A$. For concreteness we focus the discussion on future null infinity. Similar considerations apply to past null infinity with $v=t+r$ playing the role of $u$.

Future null infinity  is reached by taking the limit $r \to \infty$ with $u,\xh=$const. The abstract manifold of such limiting endpoints is denoted by $\Ip$. It has the topology of a cylinder, parameterized by $(u,\xh)$. 


We assume the Maxwell field satisfies the standard asymptotic expansion \cite{strom0}:
\ba
F_{AB}(r,u,\xh) & =&  \sum_{n=0}^{\infty} r^{-n} \Fn_{AB}(u,\xh) \\
F_{ru}(r,u,\xh) & =&  \sum_{n=2}^{\infty} r^{-n} \Fn_{ru}(u,\xh) , \label{fallrFru}
\ea
and similar expansion for the remaining components, with $F_{Au}= O(r^{0})$ and $F_{Au}= O(r^{-2})$. Since the charged scalar field is massive, we assume $F_{\mu \nu}$ satisfies the free field equations near null infinity. To solve the equations it is convenient to fix a gauge and work with the vector potential $A_\mu$. Either in radial gauge $A_r=0$ or in Lorenz gauge  one arrives at equations that can be recursively solved in terms of unconstrained, free data $\Ao_A(u,\xh)$ (plus integration constants; see e.g. \cite{lambert}). For instance, the leading $\nabla^b F_{ub}=0$ equation yields:
\be
\partial_u \Ftwo_{r u}= D^A \Fo_{Au}, \label{eqnull1}
\ee
with $ \Fo_{Au}= -\partial_u \Ao_A$.
This can be solved in terms of the free data after fixing the integration `constant'
\be
 \Ftwo_{r u}(u=-\infty,\xh) .  
\ee
For the free data we assume standard  $u \to \pm \infty$  fall-offs:
\be
\Ao_A(u,\xh)  = \Ao_A(\xh)^\pm +O(|u|^{-\epsilon}) .  \label{fallAA} 
\ee
In appendix \ref{nullapp} we show that  the field equations together with $\Ao_A(u,\xh)=O(1)$   imply
\be
\Fn_{ru}(u,\xh) =O(|u|^{n-2}) \label{falluFru}
\ee
at $u \to \pm \infty$.


The gauge parameter $\Lambda$ has a similar expansion with 
\be
\Lambda(r,u,\xh) = \lam_+(\xh) + \sum_{n=1}^{\infty} (r^{-n} \Lamn(u,\xh) + r^{-n} \ln r \Lamln(u,\xh)). \label{lamexpru}
\ee
The presence of logarithmic terms is due the $O(1)$ term \cite{subqed}. 

\subsection{Field expansion at spatial infinity} \label{spisec}
To study the fields at spatial infinity we use hyperbolic coordinates  in the $r>|t|$ region,
\be
\rho:= \sqrt{r^2 -t^2}, \quad \tau := \frac{t}{\sqrt{r^2-t^2}}
\ee
in terms of which the Minkowski line element reads:
\be
ds^2= d \rho^2 + \rho^2 d \sigma^2
\ee
with 
\be
d \sigma^2 = - \frac{d \tau^2}{1+\tau^2}+(1+\tau^2) q_{AB} dx^A dx^B = : h_{\alpha \beta} d y^\alpha d y^\beta
\ee
the (unit-radius)  three-dimensional de Sitter metric. We denote by $\H$ the abstract unit de Sitter metric that represents spatial infinity. For later convenience we introduced the notation $y^\alpha=(\tau,x^A)$ for  coordinates on $\H$. 

Following \cite{beig}, we assume the Maxwell field has a $\rho \to \infty$ expansion
\ba
F_{\alpha \beta}(\rho,y) & =&  \sum_{n=0}^{\infty} \rho^{-n} \Fn_{\alpha \beta}(y) \\
F_{\alpha \rho}(\rho,y) & =&  \sum_{n=1}^{\infty} \rho^{-n} \Fn_{\alpha \rho}(y). \label{fallrhoF}
\ea
Since the charged field is massive we assume  $F_{\mu \nu}$ satisfies the free field equations near spatial infinity. These equations 
yield evolution equations for the family of fields $\Fn_{\alpha \beta}, \Fn_{\alpha \rho}$ on $\H$.  Free field asymptotic conditions also determine the   $\tau \to \pm \infty$ behavior of these functions. For instance  (see appendix \ref{appfalltau}):
\be
  \Fn_{\t \rho}(\tau,\xh)= O(|\tau|^{n-4}), \label{fallrttauF}
\ee
a condition that will be used when relating with the fields at null infinity.

For the purposes of this paper we will only be interested in the leading equations:
\be
D^{\alpha}\Fone_{\alpha \rho}=0, \quad D_{[\alpha} \Fone_{\beta] \rho}=0, \label{eqspi1}
\ee
\be
D^\beta \Fo_{\alpha \beta}=0 , \quad D_{[\alpha}\Fo_{\beta \gamma]}=0 , \label{eqspi2}
\ee
where $D_\alpha$ is the covariant derivative on $\H$.   
Formally these equations look like electro- and magneto-static vacuum equations on $\H$, with $\Fone_{\alpha \rho}$ and  $\Fo_{\alpha \beta}$ playing the role of electric and magnetic fields respectively. They can be  solved in terms of `potentials' $\psi, \psit$ on $\H$ as follows:
The second equation in (\ref{eqspi1}) implies that $\Fone_{\alpha \rho}$ is a total derivative:
\be
\Fone_{ \alpha \rho} = D_\alpha \psi \label{Fpsi}
\ee
for some function $\psi$ on $\H$ defined up to an additive constant.\footnote{If we work with a vector potential of the form  $A_\alpha= O(\rho^0) $ and $A_\rho=O(\rho^{-1})$ then $\psi \equiv \Aone_\rho$. We will use this identification in section \ref{pssec}.} The first equation in (\ref{eqspi1}) then implies that $\psi$ satisfies the wave equation on $\H$,
\be
D^\alpha D_\alpha \psi=0. \label{wavepsi}
\ee
Equations (\ref{eqspi2}) take the same form as (\ref{eqspi1}) when written in terms of the magnetic field $\frac{1}{2}\e^{\a \b \g} \Fo_{\b \g}$ where $\e_{\a \b \g}$ is the volume form on $\H$.  They are then similarly solved in terms of a field $\psit$ such that
\be
   \Fo_{\a \b}= \e_{\a \b \g}D^\g \psit , \quad  \quad    D^\alpha D_\alpha \psit=0 . \label{eqpsit}
\ee

The gauge parameter has a similar $\rho \to \infty$ expansion
\be
\Lambda(\rho,y)= \lam(y) + O(\rho^{-1}). \label{lamspi}
\ee
Equation (\ref{eom3}) then implies 
\be
D^\alpha D_\alpha \lam =0. \label{wavelam}
\ee

The $\t \to \pm \infty$ behavior of $\lam$ is such that (see appendix \ref{waveeqdS} for further details)
\be
 \lam(\t,\xh)=\lam_\pm(\xh) + O(\t^{-\epsilon}). \label{falllamtau}
 \ee 
By performing change of coordinates between $(\rho,\tau)$ to $(r,u)$ coordinates, one can verify that $\lam_+(\xh)$ in (\ref{falllamtau}) coincides with $\lam_+(\xh)$ in (\ref{lamexpru}) (and similar statement for $\lam_-(\xh)$ at past null infinity).  See \cite{clmass} for the analogous result at time-like infinity and  appendix \ref{appfalltau} for analogous statement for the field strength.

\subsection{Field expansion at time infinity} \label{tisec}

Coordinates adapted to time-like infinity are completely analogue to the space infinity coordinates with the role of radial and time coordinates interchanged. For concreteness we focus on future time-infinity. 
In the $r<|t|, t>0$ region define:
\be
\t:= \sqrt{t^2-r^2}, \quad \rho := \frac{r}{\sqrt{t^2-r^2}}
\ee
in terms of which the Minkowski line element reads:
\be
ds^2= -d \t^2 + \t^2 d \sigma^2
\ee
with 
\be
d \sigma^2 = \frac{d \rho^2}{1+\rho^2}+\rho^2 q_{AB} dx^A dx^B = : h_{\alpha \beta} d y^\alpha d y^\beta
\ee
the unit radius hyperbolic space. We denote by $\Hp$ this abstract hyperboloid,  representing time-like infinity, with coordinates $y^\alpha=(\rho,x^A)$. Assuming free field $\t \to \infty$ fall-offs for the Maxwell field and scalar  one has:
\be
F_{\alpha \beta}(\t,y)  =  \sum_{n=0}^{\infty} \t^{-n} \Fn_{\alpha \beta}(y) , \quad F_{\alpha \t}(\t,y)  =  \sum_{n=1}^{\infty} \t^{-n} \Fn_{\alpha \rho}(y) \label{falltauF} 
\ee
\be
j_\alpha(\t,y)  = \sum_{n=3}^{\infty} \t^{-n} \jn_{\alpha}(y) , \quad j_\t(\t,y) = \sum_{n=3}^{\infty} \t^{-n} \jn_{\alpha}(y) 
\ee
To leading order, the field equations are:
\be
D^{\alpha}\Fone_{\t \alpha}=\jthree_\t, \quad D_{[\alpha} \Fone_{\beta] \t}=0. \label{eqti1}
\ee
\be
D^\beta \Fo_{\alpha \beta}=0 , \quad D_{[\alpha}\Fo_{\beta \gamma]}=0  \label{eqti0}
\ee

As in the spatial infinity case, these equations can be solved in terms of scalars $\psi, \psit$ such that
\be
\Fone_{ \alpha \t}= \partial_\alpha \psi, \quad \quad - D^\alpha D_\alpha \psi = \jthree_\t \label{pneq}
\ee
\be
   \Fo_{\a \b}= \e_{\a \b \g}D^\g \psit , \quad    D^\alpha D_\alpha \psit=0 .
\ee
They take precisely the form of electro and magneto static equations on $\Hp$ with a charge density $\jthree_\t$. Solutions to these equations are described in appendix \ref{timeapp}.

For the gauge parameter we have
\be
\Lambda(\t,y)= \lam(y) + O(\t^{-1}). \label{lamti}
\ee
with $\lam$ satisfying Laplace equation on $\Hp$
\be
D^\alpha D_\alpha \lam =0. \label{laplacelam}
\ee

\section{$\I^\pm$ conserved charges from spatial infinity perspective} \label{sec3}
In \cite{strommass} it was argued that in a general scattering process of charged particles the following identity holds:
\be
\Ftwo_{ru}(u=-\infty,\xh) = \Ftwo_{rv}(v=\infty,-\xh) \quad   \label{conslaw}
\ee
which represents an infinity of conservations laws  between past and null infinity data. If we now smear (\ref{conslaw}) with sphere functions $\lam_\pm(\xh)$ satisfying the `antipodal matching' \cite{strom0}
\be
\lam_+(\xh)= \lam_-(-\xh)  ,\label{matchlam}
\ee
 and define:
\ba
Q_+[\lambda_+] & := &\int_{S^2} d^2 V \lambda_+(\xh) \Ftwo_{ru}(u=-\infty,\xh)  \label{Qp} \\
Q_-[\lambda_-] & := & \int_{S^2} d^2 V \lambda_-(\xh)  \Ftwo_{rv}(v=\infty,\xh) \label{Qm}
\ea
the conservation law reads \cite{strom0}:
\be
Q_+[\lambda_+] = Q_-[\lambda_-]  \label{Qcons} 
\ee
for all $\lam_\pm$ satisfying (\ref{matchlam}).  The quantities  $Q_\pm[\lambda_\pm]$ have the interpretation of  large $U(1)$ gauge charges at future/past  infinities \cite{strommass}: Using Eq. (\ref{eqnull1}) one has,
\be
Q_+[\lambda_+] =   - \int_{\I^+} du d^2 V  \lambda_+(\xh) D^A \Fo_{Au}(u,\xh) + \int_{S^2} d^2 V \lambda_+(\xh) \Ftwo_{ru}(u=\infty,\xh) . \label{Qibp}
\ee
The first term in (\ref{Qibp}) is the `soft part' of the charge and depends on the Maxwell field at null infinity. The second term in (\ref{Qibp}) is the `hard part' of the charge which, upon using the field equations, depends on the massive field at time-like infinity. We refer the reader to  appendix \ref{timeapp} for further details, where we show that the hard part (and hence  $Q_+[\lambda_+]$) coincides with the expression  obtained  from covariant phase space methods \cite{clmass}. Similar considerations apply to $Q_-[\lambda_-]$.


We now show that the conservation laws (\ref{conslaw}) are a consequence of the field equations at spatial infinity. The first task is to express each side of the equality (\ref{conslaw}) in terms of fields at spatial infinity. In  appendix \ref{spinullapp} it is shown that the fall-offs (\ref{fallrFru}), (\ref{falluFru}), (\ref{fallrhoF}), (\ref{fallrttauF}) imply:
\ba
\Ftwo_{ru}(u=-\infty,\xh) &  = & \lim_{\tau \to \infty} \tau^3 \Fone_{\rho \tau}(\tau,\xh) \label{spinull1}\\
\Ftwo_{rv}(v=+\infty,\xh) &  = & -\lim_{\tau \to - \infty} \tau^3 \Fone_{\rho \tau}(\tau,\xh) .  \label{spinull2}
\ea
Recall from Eq. (\ref{Fpsi}) that $\Fone_{\alpha \rho}= D_\alpha \psi$. 
From Eq. (\ref{fallrttauF}) we then have that the $\t \to \pm \infty$ asymptotic form of $\psi$ is
\be
\psi(\t,\xh) \overset{\t \to \pm \infty}{\longrightarrow} k_\pm +  \t^{-2} \psi_\pm(\xh) + \ldots \label{psiC}
\ee
where  $k_\pm$ are possible  $O(\t^0)$ constants and the dots indicate subleading terms.  Eqns. (\ref{spinull1}), (\ref{spinull2})  then take the form:
\ba
\Ftwo_{ru}(u=-\infty,\xh) &  = & 2 \psi_+(\xh) \label{spinull1b}\\
\Ftwo_{rv}(v=+\infty,\xh) &  = & - 2 \psi_-(\xh) \label{spinull2b},
\ea
and  the conservation law  (\ref{conslaw}) translates into 
\be
\psi_+(\xh) = - \psi_-(-\xh). \label{psipm}
\ee
This equality can now be established as a consequence of the wave equation (\ref{wavepsi}) and the asymptotic form (\ref{psiC}). A proof is given in appendix \ref{waveeqdS}. 
\section{Canonical charges at spatial infinity} \label{pssec}
The  conserved charges $Q_{\pm}[\lambda_\pm]$  described in the previous section can be understood 
as large $U(1)$ gauge charges at future/past infinities \cite{strommass,clmass}. Here we show that they can also be understood as  large $U(1)$ gauge charges at spatial infinity.

We first construct the covariant phase space symplectic structure associated to constant $\t$  slices and  consider large gauge $U(1)$ symmetries at spatial infinity. We   construct the associated canonical charges $Q_\t[\lambda]$ and show that they are conserved. Their $\t \to \pm \infty$ limit is then shown to reproduce the charges   $Q_{\pm}[\lambda_\pm]$ of the previous section. In this way, the equality Eq. (\ref{Qcons}) appears as a consequence of the conservation of $Q_\t[\lambda]$.

During the analysis we will find interesting features in the asymptotics of the  symplectic structure that may be of interest in their own right. We comment on them in a final subsection.

\subsection{Symplectic current for $\t=$ const. slicing}
In the covariant phase space approach \cite{leewald,abr}, the symplectic product is defined as the flux through a Cauchy slice of a conserved symplectic current $\w^a$. For scalar QED, the standard  current is given by:
\be
\wbulk^a= \sqrt{\eta}(\delta F^{a b} \wedge \delta A_b + (\delta \D^a \phi)^* \wedge \delta \phi + c.c.)  \label{wbulk}
\ee
(the wedge is in field space). The  subscript `bulk' is to indicate that the current may have to be amended by a total-derivative `boundary' term. A way to determine whether such boundary term is required is  to compute the `leakage' of the symplectic current between two Cauchy slices \cite{ih}. Only for zero leakage is the symplectic product well-defined (i.e. independent of the Cauchy slice). In \cite{ih} this strategy is followed to obtain the boundary contribution from a horizon boundary. Here we follow the same strategy with spatial infinity as a boundary. 

To proceed we need to specify fall-off conditions on the vector potential $A_\mu$. We assume a power series expansion in $\rho^{-1}$ with leading terms given by:
\be
A_\alpha= O(\rho^0) , \quad A_\rho=O(\rho^{-1}) \label{fallA}
\ee
which is compatible with the fall-offs of $F_{\mu \nu}$ described in section \ref{spisec}.
The leakage of the current (\ref{wbulk}) between two Cauchy slices $\t=\t_1$ and $\t=\t_2$  is then given by:
\be
\lim_{\rho \to \infty} \int_{\t_1}^{\t_2}d \t \int d^2 \xh \;  \wbulk^\rho =   \int_{\Delta} d^3 V h^{\alpha \beta}\delta \Fone_{\rho \alpha} \wedge \delta \Ao_\beta 
\ee
where $\Delta \subset \H$ is the region bounded by the spheres $\t=\t_1,\t_2$ on $\H$. 
Written in terms of the vector potential, the leaking term reads:
\ba
 h^{\alpha \beta}\delta \Fone_{ \rho \alpha} \wedge \delta \Ao_\beta & =& - \delta D^\alpha \Aone_\rho \wedge \delta \Ao_\alpha \\
 & =& -D^\alpha(  \delta \Aone_\rho \wedge \delta \Aoa) + \delta \Aone_\rho \wedge \delta D^\alpha \Aoa  \label{leak}
\ea 
(indices in $\H$ are raised with $h^{\alpha \beta}$). Since the first term in (\ref{leak}) is  a total derivative, it can be cancelled by adding a boundary term to (\ref{wbulk}). The appropriate boundary term is
\be
\wbdy^a := \partial_b( \sqrt{\eta}\, \delta A^a \wedge \delta A^b). \label{wbdy}
\ee
The second term in (\ref{leak}) however remains. This term can be eliminated by imposing appropriate gauge condition on the leading vector potential. For instance any gauge condition of the form
\be
\Aone_\rho  \propto D^\alpha \Ao_\alpha  \label{gralgge}
\ee
will make the unwanted contribution to vanish. In fact, our interest is in studying large $U(1)$ gauge parameter in Lorenz gauge. To leading  order in $\rho \to \infty$ this gauge condition implies
\be
2 \Aone_\rho + D^\alpha \Ao_\alpha =0, \label{gaugecond}
\ee
which is a particular case of (\ref{gralgge}). 
To summarize: The symplectic structure associated to the $\t=$ const. foliation, with fall-off conditions (\ref{fallA}) and gauge condition (\ref{gralgge}) is given by:
\be
\Omega= \int_{\Sigma_\t} dS_a (\wbulk^a+\wbdy^a) \label{Omega}
\ee
with $\wbulk^a$ and $\wbdy^a$ given in (\ref{wbulk}) and (\ref{wbdy}) respectively.\footnote{The fall-offs (\ref{fallA}) imply a logarithmically divergent term in the $\rho$ integral of (\ref{Omega}). This would-be divergence however can be shown to be zero due to properties satisfied by  the fields on $\H$. This and related points are discussed in subsection \ref{sympsec}. 
\label{fnote1}}

\subsection{Large $U(1)$ gauge transformations and associated charges}
We now consider large gauge transformations that are nontrivial at spatial infinity, 
\be
\Lam(\rho,y) =\lam(y) +O(\rho^{-1}). \label{falllam}
\ee
The associated  charges $Q[\lambda]$ are defined by the condition\footnote{The charge will depend on $\Lambda$ only through its asymptotic value $\lambda$ hence the notation $Q[\lambda]$.}
\be
\delta Q[\lambda]= \Omega(\delta_\Lambda,\delta) \label{deltaQ}
\ee
where $\delta_\Lambda$ is the action of the gauge transformation, Eq. (\ref{gge}). The contribution to (\ref{deltaQ}) due to the `bulk' symplectic current becomes, upon using the field equations, the  total derivative term
\be
\wbulk^a(\delta_\Lambda,\delta)= -\delta \partial_b(\sqrt{\eta} \Lambda F^{ab}).
\ee
On the other hand, the contribution from $\wbdy^a$ is a total derivative to begin with:
\be
\wbdy^a(\delta_\Lambda,\delta)= \delta \partial_b \sqrt{\eta}(\nabla^a \Lambda A^b - A^a \nabla^b \Lambda).
\ee
The integral over $\Sigma_\t$ thus becomes a  $\rho \to \infty$ surface integral. The fall-offs (\ref{fallA}), (\ref{falllam}) imply the charge is given by:
\be
Q_\t[\lambda] = \int_{C_\t} dS_\alpha \sqrt{h} h^{\alpha \beta}(\partial_\beta \lam \Aone_\rho - \lam \Fone_{\beta \rho}), \label{Qt1}
\ee
where $C_\t$ is the $\t=$const. surface on $\H$. For the fall-offs (\ref{fallA}) we have $\Fone_{\alpha \rho} = \partial_\alpha \Aone_\rho$. Comparing with (\ref{Fpsi}) we see that $\Aone_\rho$ is the field $\psi$ of the previous sections. Thus (\ref{Qt1}) can be written as:\footnote{An analogue charge is given in \cite{compere} for supertranslations in gravity. Our argument for charge conservation is the same as the one given there.}
\be
Q_\t[\lambda] =  \int_{C_\t} dS_\alpha \sqrt{h} h^{\alpha \beta}(\partial_\beta \lam \psi - \lam \partial_\beta \psi). \label{Qt2} \\
\ee
This has the form of a symplectic product of two scalar fields $\psi$ and $\lam$ on $\H$. 
In section \ref{spisec} we saw that both $\psi$ and $\lam$ satisfy the wave equation on $\H$,  (\ref{wavepsi}) and (\ref{wavelam}) (the latter arising from the Lorenz gauge condition). 
It thus follows that (\ref{Qt2}) is independent of the slice $C_\t$. That is,  the charges $Q_\t[\lambda]$ are conserved.  Finally, it is easy to verify that the $\t \to \infty$ fall-offs (\ref{falllamtau}), (\ref{psiC})  imply
\be
\lim_{\t \to \pm \infty}Q_\t[\lambda] = Q_{\pm}[\lambda_\pm]
\ee
with $Q_{\pm}[\lambda_\pm]$ the charges defined in Eqns. (\ref{Qp}), (\ref{Qm}) (just evaluate (\ref{Qt1}) for $\alpha=\t$ and use $\sqrt{h} h^{\t \t}= - (1+\t^2)^{3/2}$). The matching condition (\ref{matchlam}) can be seen as a consequence of the wave equation and fall-offs on $\lambda$, see appendix \ref{waveeqdS}.

\subsection{Asymptotic properties of the symplectic structure} \label{sympsec}
 Let us denote by $\Okg_\t(\delta,\delta')$ the Klein-Gordon (KG)  symplectic product on $\H$ so that  Eq. (\ref{Qt2})  takes the form
 \be
Q_\t[\lambda] = \Okg_\t(\lam,\psi).
 \ee
 
As discussed in appendix \ref{waveeqdS}, the fields $\lam$ and $\psi$ have different $\t \to \pm \infty$ fall-off behavior, associated to different `free data' that can be prescribed in the asymptotic  boundary of $\H$. This in turn implies that the KG symplectic product between $\lam$'s or $\psi$'s fields vanish
\be
 \Okg_\t(\lam,\lam')=   \Okg_\t(\psi,\psi')=0 \label{okgzero}.
\ee
 In other words, the phase space of massless KG fields on $\H$ has a ``$(q,p)$'' decomposition with $\lam$-type fields playing the role of $q$ and $\psi$-type fields the role of $p$.\footnote{From the point of view of the $\t=0$ slice, this decomposition is related  to `parity conditions'  analogous to those in gravity \cite{rt}. See appendix B.2.C of \cite{mizera} for a discussion in the context of massless scalars.} There are two interesting observations that follow.
 
The first one is in relation to the boundary phase space of gauge theories given in \cite{freidel}. The authors consider gauge theories in finite spatial regions. In order to have good gluing properties of different regions, the authors are lead to introduce a  boundary phase space consisting of a conjugate pair of normal electric field and gauge phase. Here we see the same type of boundary phase space: If  $\lam$ is regarded as a dynamical variable, it follows that $\lam$ and $\psi$ are canonically conjugated (with $Q_\t$   expressing their symplectic product).  In the language of  \cite{freidel}, $\psi$ plays the role of normal electric field and $\lam$ is the conjugated  phase.

The second observation is regarding a potential logarithmic divergence in the symplectic product (\ref{Omega}). Expressing the integral over $\Sigma_\t$ in (\ref{Omega}) as $\lim_{R \to \infty} \int^R_0 d \rho \int d^2 \xh ( \ldots )$ one finds
\be
\Omega_\t(\delta,\delta') = \lim_{R \to \infty}  \log(R) \; \Okg_\t(\delta \psi,\delta' \psi) +  \text{finite}.
\ee
However, by virtue of (\ref{okgzero}) this would-be divergent piece vanishes.\footnote{To simplify the discussion we have omitted a second  logarithmic divergence proportional to  \mbox{$\sim  \int dS_\alpha \sqrt{h} h^{\alpha \beta} \delta \Fo_{\alpha \beta} \wedge \delta\Ao_\beta$}. This can also be shown to be zero by entirely similar arguments.}


\section{Magnetic charges} \label{magsec}
Dual `magnetic' charges can be obtained by considering the previous charges with  
\be
F^*_{ab} := \frac{1}{2}  \epsilon_{abcd}F^{cd}
\ee
playing the role of $F_{ab}$. From the null infinity perspective we have
\be
\Fstwo_{ru}= \frac{1}{2}\e^{AB} \Fo_{AB} \label{Fstwo}
\ee
and similarly for $\I^-$. The magnetic versions of the conservation law (\ref{conslaw}) and charges (\ref{Qp}), (\ref{Qm}) are defined by the replacement $F_{ab} \to F^*_{ab}$. For instance, we have
\be
\Qt_+[\lambda]  =  - \int_{\I^+} du d^2 V  \lambda(\xh) \partial_u \Fstwo_{ru}(u,\xh) + \int_{S^2} d^2 V \lambda(\xh)\Fstwo_{ru} (u=\infty,\xh) \label{Qtibp}
\ee
as the magnetic analogue of (\ref{Qibp}). Using (\ref{Fstwo}) we see that the first term in (\ref{Qtibp}) is the magnetic `soft' charge that features in Weinberg's soft theorem \cite{subqed}.  In appendix \ref{timeapp} we show that the field equations at time-like infinity imply that the second term in (\ref{Qtibp}) vanishes and there is no `hard' contribution to the magnetic charge.

At spatial infinity, the starred field strength is given by
\ba
\Fso_{\alpha \beta} & = & \e_{\alpha \beta}^{\phantom{\a \b} \gamma}  \Fone_{\gamma \rho} \\
\Fsone_{\alpha \rho} & = & \frac{1}{2} \e_{\alpha}^{\phantom{\a} \beta \gamma} \Fo_{\b \g}.
\ea
The definition of $\psit$ in (\ref{eqpsit}) then implies
\be
\Fsone_{ \alpha \rho} = D_\alpha \psit.
\ee
Thus, the replacement $F_{\mu \nu} \to F^*_{\mu \nu}$  corresponds to $\psi \to \psit$ (and $\psit \to - \psi$). In particular, the conserved magnetic charges at spatial infinity are
\be
\Qt_\t[\lambda] = \int_{C_\t} dS_\alpha \sqrt{h} h^{\alpha \beta}(\partial_\alpha \lam \psit - \lam \partial_\alpha \psit). \label{Qtt2}
\ee
As for  the electric charges, they satisfy $\lim_{\t \to \pm \infty} \Qt_{\t}[\lambda]= \Qt_{\pm}[\lambda_\pm]$.

\section{Outlook} \label{finalsec}

In this paper we studied Maxwell equations at spatial infinity in order to shed light into the ``conservation laws'' (\ref{newcons}) that have recently been shown to be implied by Weinberg's soft photon theorem. It has long been known \cite{hansen,held} that an appropriate  way to think of spatial infinity is as a unit time-like hyperboloid $\H$  parametrizing all asymptotic directions of spatial geodesics. Information at the future end of $\I^-$ is propagated along $\H$ and reaches the past end of $\I^+$. In this way, condition (\ref{newcons}) appears as a consequence of the field equations at $\H$.

We then set to explore large $U(1)$ gauge symmetries and charges from the perspective of $\H$.  At future/past infinities these charges are known to be given by a smeared version of (\ref{newcons}). In order to get  `finite time' versions of these charges, we considered a spacetime foliation of hypersurfaces $\Sigma_\t$ that intersect $\H$ on constant $\t$ spheres. After imposing an asymptotic Lorenz gauge condition, we  obtained the canonical charge  $Q_\t[\lambda]$ associated to a   large gauge symmetry $\lambda$. The properties of the field equations at spatial infinity imply that  $Q_\t[\lambda]$ is conserved and its $\t \to \pm \infty$  limit given by the known future/past infinity charges. 

While deriving the charges  we encountered several interesting properties of the symplectic structure that may be of interest in their own: The need to include a boundary term, the relation with the boundary phase space of \cite{freidel}, and the absence of logarithmic divergences  as a consequence of the properties of the fields on $\H$.

Many future directions appear in sight. Extensions of the ideas presented here to (at least perturbative) gravity seem within reach.  Indeed many of the ingredients can already be found in the literature: The supertranslation charges at spatially infinity (in an asymptotic harmonic-like gauge)  given in \cite{compere} are completely analogue to the charges $Q_\t[\lambda]$ found here. The extensions of supertranslations and superrotations from null to time-like infinity described in \cite{clbmsmass} have a natural counterpart in the analogue problem of going from null to space-infinity.  Finally, it would be interesting to explore possible connections to  studies in Minkowski holography \cite{solodukhin,marolf,sundrum,stromholo}. \\ 

\noindent {\bf Acknowledgements}\\
We are greatly indebt to Abhay Ashtekar and Alok Laddha for key discussions. MC would like to thank the participants of the Perimeter Institute workshop `Infrared problems in QED and quantum gravity', specially to Laurant Freidel, for stimulating  discussions. 

\appendix
\section{More on fall-offs and asymptotic field equations} \label{appA}
\subsection{Eq. (\ref{falluFru}) } \label{nullapp}
A convenient gauge to solve the free Maxwell equations is radial gauge $A_r=0$ (see  \cite{lambert} for a complete discussion). Here we only discus the equations needed to determine the $u \to \infty$ asymptotic behavior given in Eq. (\ref{falluFru}). Assuming a $r^{-n}$ power expansion for the vector potential, the equation $\nabla_b F^{Ab}=0$ yields 
\be
\partial_u \An_A = \frac{1}{2n} D^B \overset{-(n-1)}{F_{AB}}, \quad \quad n \geq 1,
\ee
which can be integrated to give $\An_A$ in terms of $\overset{-(n-1)}{A_A}$. If  $\Ao_A=O(|u|^0)$ as in Eq. (\ref{fallAA}) this  implies
\be
\An_A=O(|u|^n) .\label{AAu}
\ee
Next, we look at the equation $\nabla_b F^{ub}=0$. The radial gauge implies this equation is equivalent to 
\be
F_{ru} = r^{-2}(k(u,\xh) + D^A A_A) \label{radggeeq1}
\ee
where $k(u,\xh)$ is an integration `constant' which for consistency we assume to be $O(|u|^0)$. Expanding this equation in $1/r$ and using  Eq. (\ref{AAu}) one concludes the asymptotic behavior of $\Fn_{ru}$ is given by Eq. (\ref{falluFru}). 

\subsection{Eq. (\ref{fallrttauF})} \label{appfalltau}
Free field Maxwell equations at spatial infinity take simple form when written  in terms of the vector potential in Lorenz gauge. For $\An$ one finds the equation
\be
D^\alpha D_\alpha \An_\rho +(n-1)(n-3) \An_\rho =0
\ee
this imply that $\An_\rho$ can be $O(\t^{1-n})$ or $O(\t^{n-3})$ at $\t \to \pm \infty$, which corresponds to $O(\t^{-n})$ or $O(\t^{n-4})$ behavior of $\Fn_{\rho \t}$.  For $n \geq 2$ the leading behavior is $O(\t^{n-4})$ which corresponds to Eq. (\ref{fallrttauF}). However for $n=1$ the leading piece is the $O(\t^{-n})$ one. One could say that the demand of this leading piece to vanish corresponds to condition of  having well-defined charges, i.e. only then Eqns. (\ref{spinull1}), (\ref{spinull2}) make sense. We now provide an argument as to how this condition arises from the assumed behavior at null infinity.

In the previous subsection we saw that $\Fk_{ru}=O(|u|^{k-2})$ at $u \to \infty$. Let us assume a $1/u$ power expansion at $u \to \infty$ for the subleading terms so that
\be
F_{ru}(r,u,\xh)= \sum_{k=2}^{\infty} \frac{u^{k-2}}{r^k} \sum_{l=0}^\infty \frac{1}{u^l} (\Fk_{ru})^{(l)}(\xh). \label{Frukl}
\ee
Consider now the above expression in the $(\rho,\t)$ coordinates adapted to spatial infinity. Substituting
\ba
r &=& \rho \sqrt{1+\t^2} = \rho \t (1+ O(\t^{-2})) ,  \label{covrurhot1} \\
u &=& \rho(\t-\sqrt{1+\t^2}) =  - \frac{\rho}{2 \t} (1+ O(\t^{-2}))\label{covrurhot2}
\ea
in (\ref{Frukl}) and using the change of coordinates relation
\be
F_{\rho \t} = \frac{\rho^2}{r}F_{ru} = \frac{\rho}{\t} (1+O(\t^{-2}))  F_{ru}
\ee
one finds (after reorganizing the sums) that indeed the leading $1/\t$ power of each $1/\rho$ term corresponds to $\Fn_{\rho \t}=O(\t^{n-4})$.


\subsection{Eqns. (\ref{spinull1}), (\ref{spinull2})} \label{spinullapp}
The equality we wish to study deals with a priori  different limits. The RHS of Eq. (\ref{spinull1}) corresponds to:
\be
 \lim_{\t \to \infty} \t^3 [\lim_{\rho \to \infty} \rho F_{\rho \t}(\rho,\t,\xh)] =: (\Fone_{\rho \t})^{(-3)}(\xh) \label{limitspi}
 \ee
whereas the LHS is:
 \be
  \lim_{u \to - \infty}[\lim_{r \to \infty} r^2 F_{ru}(r,u,\xh)] =: (\Ftwo_{ru})^{(0)}(\xh) \label{limitnull}
\ee
The idea is to capture the above, a priori different, limits as a single limit along an appropriately chosen curve $\gamma(s)$ in spacetime,
\be 
s \mapsto \gamma(s)=(t(s),r(s),\xh). 
\ee
A curve that will do the jobs is: 
\be
\rho(s) = s^2, \quad \t(s)=s,\label{rhotcurve}
\ee
whose form in  $(r,u)$ coordinates can be obtained from (\ref{covrurhot1}), (\ref{covrurhot2}):
\ba
r(s) = s^3+O(s) , \quad u(s) = - \frac{s}{2}  +O(1/s) .\label{rucurve}
\ea
The intuitive picture of this curve is as follows. Eq. (\ref{limitspi}) tell us to first take $\rho \to \infty$ before taking  $\t \to \infty$. The curve (\ref{rhotcurve}) accomplishes this goal by making $\rho(s)$ diverge faster than $\t(s)$ as $s \to \infty$. Similarly Eq. (\ref{limitnull}) tell us that `$r$ should go to infinity faster than $u$ goes to (minus) infinity'. This  is precisely the $s \to \infty$ behavior of the curve in $(r,u)$ coordinates (\ref{rucurve}). 

The fall-offs described in section \ref{prelsec} ensure this intuitive picture is correct and that one can replace the limits in Eqns. (\ref{limitspi}), (\ref{limitnull}) by a single $s \to \infty$ limit. For (\ref{limitspi}), it is easy to verify that the expansion (\ref{fallrhoF}) together with $\t$ fall-off (\ref{fallrttauF}) imply
\be
\lim_{s \to \infty} s^5 F_{\rho \t}(x=\gamma(s)) = (\Fone_{\rho \t})^{(-3)}(\xh) . \label{limFrhotau}
\ee
Similarly, it is easy to verify that the $r \to \infty$ expansion (\ref{fallrFru}) together with the $u$ fall-offs (\ref{falluFru}) imply
\be
\lim_{s \to \infty} s^6 F_{r u}(x=\gamma(s)) = (\Ftwo_{ru})^{(0)}(\xh) . \label{limFru}
\ee
Finally, the relation between $F_{\rho \t}$ and $F_{ru}$ from the change of coordinates, $F_{\rho \t} = \frac{\rho^2}{r}F_{ru}$, implies
\be
 F_{\rho \t} = (s + O(1/s))F_{ru} \label{Frhotru}
 \ee
  along the $\gamma(s)$ curve. Substituting  (\ref{Frhotru}) in (\ref{limFrhotau}) we see that this limit is precisely the limit (\ref{limFru}). It then follows that  $ (\Fone_{\rho \t})^{(-3)}(\xh)= (\Ftwo_{ru})^{(0)}(\xh)$. This establishes Eq. (\ref{spinull1}). Eq. (\ref{spinull2}) can be shown by an analogous argument.

\section{Wave equation on $\H$}  \label{waveeqdS}
\subsection{Green's functions} 
In this section we describe the solutions to the wave equation 
\be
D^\alpha D_\alpha \psi=0  \label{waveeq}
\ee
in terms of asymptotic data at one asymptotic boundary of $\H$. For definitiveness we consider data at the past asymptotic boundary, $\tau \to - \infty$ but similar considerations apply to future asymptotic boundary data. By analyzing the wave equation (\ref{waveeq}) in the $\tau \to - \infty$ limit, one concludes there are two types of asymptotic solutions:
\ba
\lambda(\t,\xh) & = & \lambda_-(\xh) + O(\t^{-2} \ln |\t|)  \label{type1} \\
\psi(\t,\xh) & = & \t^{-2} \psi_-(\xh) +O(\t^{-4}) \label{type2} 
\ea
The leading terms   $ \lambda_-(\xh)$, $ \psi_-(\xh) $ are unconstrained  and represent free data in terms of which subleading terms are determined. As  indicated in the notation, the first type of solution is  relevant  for large gauge parameters  whereas the second type is relevant for the leading component of the field strength.\footnote{Recall $\psi$ in Eq. (\ref{Fpsi}) is defined up to an overall constant. For the following argument we choose such constant so that $k_-=0$ in Eq. (\ref{psiC}).}  We wish to  obtain integral expressions for the solutions $\lambda(y)$, $\psi(y)$ in terms of the corresponding asymptotic data:
\ba
\lambda(y) &=& \int d^2  V' \; G^{(0)}(y; \xh') \lambda_-(\xh') \label{intlam} \\
\psi(y) &= & \int d^2 V' \; G^{(2)}(y; \xh') \psi_{-}(\xh') ,
\ea
where the $G$'s are appropriate Green's functions.  In the following we describe these Green's functions. For that purpose let us introduce some notation. Given  $y \in \H$ and $\xh' \in S^2$ define:
\ba
Y^\mu & :=& ( \tau, \sqrt{1+\t^2} \; \xh)\label{defY} \\
\sigma &:=&  \t+ \sqrt{1+\t^2} \, \xh \cdot \xh' ,\label{defsigma}
\ea
that is,  $y \in \H$ defines a point  $Y^\mu$ in the unit hyperboloid of Minkowski space. $\xh'$ defines the point on the past light cone: $(-1,\xh')^\mu$.  $\sigma$ is then the inner product between these two vectors.

We now study each Green's function separately.

\subsubsection{Green's function for Eq. (\ref{type1})}
The Green's function in (\ref{intlam}) is a Lorentzian version of the Green's function used in \cite{clmass} to obtain the gauge parameter at time-like infinity.\footnote{The Green's functions of both signatures are particular cases of the  more general discussion given in \cite{solodukhin}.} Taking the appropriate analytic continuation of the expression given in \cite{clmass} yields:
\be
G^{(0)}(y; \xh') = - \frac{1}{4 \pi \sigma^2}. \label{G0}
\ee
Let us verify that indeed (\ref{G0}) satisfies the desired requirements. The action of the wave operator $D^\alpha D_\alpha$ on a function $f(\sigma)$  can be easily calculated by expressing $D^\alpha D_\alpha$  in terms of the Minkowski space d'Alembertian (see \cite{cgreen} for similar treatment at time-like infinity).  One obtains:
\be
D^\alpha D_\alpha f(\sigma) = - \sigma^2 f''(\sigma) - 3 \sigma f'(\sigma). \label{boxfsigma}
\ee
It then follows that  $f(\sigma)= \sigma^{-2}$ and hence $G^{(0)}(y; \xh') $ satisfies the wave equation. We now verify the boundary condition. The $\t \to -\infty$ behavior of   $\sigma$ is
\be
 \sigma  \overset{\t \to - \infty}{\longrightarrow}  \t(1 - \xh \cdot \xh')  - \frac{\xh \cdot \xh'}{2 \t} + \ldots
\ee
From which we have
\be
G^{(0)}(y; \xh') \overset{\t \to - \infty}{\longrightarrow} \left\{ \begin{array}{lll} O(\t^{-2}) & \text{if} & \xh \neq \xh' \\ O(\t^{2}) & \text{if} & \xh = \xh'  \end{array} \right. 
\ee
On the other hand one can verify that $\int d^2 V' G^{(0)}(y; \xh') = 1$. It then follows that $G^{(0)}(y; \xh') \overset{\t \to - \infty}{\longrightarrow} \delta^{(2)}(\xh,\xh')$ which then  imposes the boundary condition (\ref{type1}). The solution given by (\ref{intlam}), (\ref{G0}) allow us to explore the $\t \to +\infty$ limiting value. In this case
\be
 \sigma  \overset{\t \to + \infty}{\longrightarrow}  \t(1 + \xh \cdot \xh')  + \frac{\xh \cdot \xh'}{2 \t} + \ldots
\ee
which now implies $G^{(0)}(y; \xh') \overset{\t \to - \infty}{\longrightarrow} \delta^{(2)}(\xh,-\xh')$. We thus recover the antipodal identification (\ref{matchlam}) of the gauge parameter at plus and minus infinity.
\subsubsection{Green's function for Eq. (\ref{type2})}
To express $\psi$ in terms of given asymptotic data, we consider the Kirchhoff integral representation (see for instance section 4.3 of \cite{poisson}):
\be
\psi(y) =  -\int_C dS'_\alpha \sqrt{h}h^{\alpha \beta}(y')\big(G_R(y,y') \partial'_\beta \psi(y') -\partial'_\beta G_R(y,y')  \psi(y') \big)  \label{kirch}
\ee
where $C$ is a Cauchy slice of $\H$ and $G_R(y,y')$ the retarded Green's function. The idea is  to consider (\ref{kirch}) with $C$  a $\tau'=$ const. surface and take the limit $\tau' \to -\infty$. In such case Eq. (\ref{kirch}) takes the form:
\be
\psi(y) = \lim_{\t' \to - \infty}  |\t'|^3 \int_{\t'=\text{const.}} d^2 V' \big(G_R(y,y') \partial_{\t'}\psi(y')  -  \partial_{\t'}G_R(y,y')   \psi(y')\big). \label{kirchinf}
\ee
To proceed we need the expression for $G_R(y,y')$. For $y$ inside the future cone of $y'$, it has to satisfy the wave equation (\ref{waveeq}) and be symmetric with respect to the isotropy group of $y'$. The solution is easily found in terms of 
\be
P=Y^\mu Y'_\mu
\ee 
 (see e.g. \cite{stromdS}) with $Y^\mu$ and $Y'^\mu$  defined in (\ref{defY}) for $y$ and $y'$ respectively. One finds:
\be
G_R(y,y') =\frac{1}{2 \pi} \theta(\t-\t') \theta(P-1)   \frac{P}{\sqrt{P^2-1}}  \label{GR} 
\ee
where $\theta$ is the step function that imposes  $y$ to be inside the cone of $y'$ and $\t>\t'$. It is easy to verify that for $y \to y'$ (\ref{GR})  approaches  the flat space retarded Green's function. It now remains to study the  $\t' \to -\infty$ behaviour of $G_R(y,y')$ in order to evaluate (\ref{kirchinf}). 
When $\t' \to -\infty$ one has
\be
P= - \t' \sigma +O(\t'^{-1}),
\ee
and
\be
G_R(y,y') = \theta(\sigma)  + O(\t^{-1}) \label{GRtau} 
\ee
with $\sigma$ as defined in (\ref{defsigma}). 
Using (\ref{type2}) and (\ref{GRtau}) in Eq. (\ref{kirchinf}) one obtains:
\be
\psi(y) = \frac{1}{\pi} \int d^2 V' \theta (\sigma) \psi_-(\xh'). \label{solpsio}
\ee 
It is instructive to verify that (\ref{solpsio}) is indeed a solution to (\ref{waveeq}) with asymptotic condition (\ref{type1}). 
First notice that, as in the $\lambda$  case, the dependance on $y$ is through $\sigma$. We can thus use (\ref{boxfsigma}) to evaluate the wave operator on (\ref{solpsi}):
\be
D^\alpha D_\alpha \psi(y) = - \frac{1}{\pi} \int d^2 V'  \left( \sigma^2 \delta'(\sigma) + 3 \sigma \delta(\sigma) \right) \psi_-(\xh') \label{boxpsi}
\ee
with $\delta= \theta'$ the delta function. For regular data $\psi_-(\xh')$ each term in (\ref{boxpsi}) vanishes. 

We now study the asymptotic behavior of $\psi(y)$. Eq. (\ref{solpsi}) tells one is integrating $ \psi_-(\xh')$ over the portion of the $\xh'$ sphere given by
\be
\xh \cdot \xh' > - \frac{\t}{\sqrt{1+\t^2}}. \label{region}
\ee
When $\t \to - \infty$, the region (\ref{region}) becomes a disk of radius $|\t|^{-1}$ centered at $\xh'=\xh$. Thus (\ref{solpsi}) becomes $\psi(y)\approx \frac{1}{\pi}( \pi \, \t^{-2} \psi_-(\xh) )$ and we recover the boundary condition (\ref{type2}). 

At the other extreme, when $\t \to \infty$ the integration region  becomes the whole sphere minus a disk of radius $\t^{-1}$ centered at $\xh'=-\xh$. Thus, in this limit the solution becomes
\be
\psi(y)  \overset{\t \to + \infty}{\longrightarrow} = \frac{1}{\pi}\big( \int_{S^2} d^2 V' \psi_-(\xh') - \pi \t^{-2} \psi_-(-\xh) \big),
\ee
which is if of the form (\ref{psiC}) with a nonzero $k_+$ and $\psi_{+}(\xh)= -\psi_-(-\xh)$. This establishes the relation (\ref{psipm}).

\section{Laplace and Poisson equations on $\Hp$} \label{timeapp}
\subsection{Solutions to leading order equations}
In this section we study solutions to Laplace's and Poisson's equations on $\H$:
\be
D^\alpha D_\alpha \psit=0 , \label{laplace}
\ee
\be
- D^\alpha D_\alpha \psi = \jthree_\t . \label{poisson}
\ee
By similar arguments  to those  given in section \ref{appfalltau}  one can show that the $\rho \to \infty$ behavior of $\psi,\psit$ is  (up to arbitrary additive constants that we set to zero) 
\be
\psi, \psit = O(\rho^{-2}) \label{psiO}.
\ee
On the other hand, Laplace's equation (\ref{laplace}) on $\H$ imply a $O(\rho^0)$ or $O(\rho^{-2})$ behavior at $\rho \to \infty$. The second kind of behavior however yields solutions that are singular at $\rho =0$ (this can for instance be established by looking at the solutions with given spherical harmonic angular dependance). Thus Eqns. (\ref{laplace}) and (\ref{psiO}) imply 
\be
\psit=0. \label{psitzero}
\ee
We not turn to Poisson's equation (\ref{poisson}). The solution for $\psi$ can be written as 
\be
\psi(y) = \int d^3 V' \G(y ; y')  \jthree_\t(y'), \label{solpsi}
\ee
with $\G(y;y')$ the appropriate Green's function. This  Green's function  can be found by similar methods as those leading to Eq. (\ref{GR}) resulting in,
\be
\G(y;y') = \frac{1}{4 \pi} \big(\frac{Y \cdot Y'}{\sqrt{(Y \cdot Y')^2-1}} -1 \big)\label{greenpn}
\ee
where $Y^\mu$ and $Y'^\mu$ are Minkowski vectors associated to $y, y' \in \Hp$, i.e.
\be
Y^\mu= (\sqrt{1+\rho^2}, \rho \xh)
\ee
and $Y \cdot Y'$ their inner product with $(+,-,-,-)$ sign convention. One can verify that (\ref{greenpn}) approaches the standard flat-space Green's function in the limit $y \to y'$.  The ``$-1$'' in (\ref{greenpn}) ensures the possible additive constant on $\psi$ is set to zero.

The solution to Laplace's equation for $\lam$ (\ref{laplacelam}) with given boundary value $\lam_+(\xh)$ was discussed in \cite{clmass}. It is given by
\be
\lam(y)= \int d^2 V' G(y;\xh') \lambda_+(\xh') , \label{loti}
\ee
where the  Green's function is 
\be
G(y;\xh')=\frac{1}{4 \pi \sigma(y,\xh')^2} \label{greentimelam}
\ee
with
\be
\sigma(y,\xh')= \sqrt{1+\rho^2}- \rho \xh \cdot \xh'
\ee
 the inner product between $Y^\mu$ and $(1,\xh')^\mu$.
\subsection{Time infinity contribution to charges} \label{poissoneq}

In this section we evaluate the  boundary  terms in Eqns. (\ref{Qibp}), (\ref{Qtibp}):
\ba
\Qp[\lambda_+] & := &\int_{S^2} d^2 V \lambda_+(\xh) \Ftwo_{ru}(u=\infty,\xh) ,\\
\Qtp[\lambda_+] & := &  \frac{1}{2} \int_{S^2} d^2 V \lambda_+(\xh) \e^{AB} \Fo_{AB} (u=\infty,\xh) .
\ea

 By a similar argument to the one given in section \ref{spinullapp}, one finds:
\ba
\Ftwo_{ru}(u=\infty,\xh) & =& \lim_{\rho \to \infty} \rho^3 \Fone_{\rho \t}(\rho,\xh) , \label{Fruti} \\
\frac{1}{2}\e^{AB}\Fo_{AB}(u=\infty,\xh) &= & \lim_{\rho \to \infty} \rho^3 \Fsone_{\rho \t}(\r,\xh). 
\ea
Recalling the relation with $\psi,\psit$, and using (\ref{psiO}) this allow us to write the time-infinity charges as
\ba
\Qp[\lambda_+] & = & \lim_{\rho \to \infty}  -2 \rho^2 \int_{S^2} d^2 V \lambda_+(\xh) \psi(\rho,\xh) , \label{Qp2} \\
\Qtp[\lambda_+] & = &  \lim_{\rho \to \infty} -2 \rho^2 \int_{S^2} d^2 V \lambda_+(\xh)\psit(\rho,\xh).
\ea
From (\ref{psitzero}) we immediately conclude that $\Qtp[\lambda_+]=0$. We now show that $\Qp[\lambda_+]$ coincides 
with the time-infinity charge obtained in \cite{clmass} by covariant phase space methods. The expression given in \cite{clmass} is
\be
\Qh[\lambda] = - \int_{\Hp} d^3 V \lam(y) \jthree_\t(y), \label{Qhti}
\ee
with $\lam$ the gauge parameter at $\Hp$  given in Eq. (\ref{loti}).  To see that (\ref{Qp2}) and (\ref{Qhti}) coincide, we consider the $\rho \to \infty$ limit of $\psi(y)$ as given in (\ref{solpsi}). In this limit
\be
Y \cdot Y'=  \rho \, \sigma(y',\xh) + O(\rho^{-1}),
\ee
\be
\frac{Y \cdot Y'}{\sqrt{(Y \cdot Y')^2-1}}-1 = \frac{1}{2 \rho^2\sigma(y',\xh)^2} +O(\rho^{-4}),
\ee
and so
\be
\G(y;y') =\frac{1}{2 \rho^{2}} G(y';\xh) + O(\rho^{-4}) \label{Gexp}
\ee
where $\G$ and $G$ are the Green's function given in Eqns. (\ref{greenpn}) and (\ref{greentimelam}) respectively.
Using the expansion (\ref{Gexp}) in (\ref{Qp2}), (\ref{solpsi}) one finds:
\ba
\Qp[\lambda_+] & = & - \int d^2 V \lambda_+(\xh) \int d^3 V' G(y';\xh) \jthree_\t(y') \\
& = & - \int d^3 V' \lam(y')  \jthree_\t(y')
\ea
where in the second line we used the expression (\ref{loti})  for the gauge parameter on $\Hp$. We  thus conclude that  $\Qp[\lambda_+]$ coincides with $\Qh[\lambda] $  given in \cite{clmass}.

\end{document}